\documentclass[twocolumn, prd, showpacs,
floats]{revtex4}
\usepackage{graphicx}
\usepackage{dcolumn}
\usepackage{bm}
\usepackage{amsmath}

\begin{document}

\title{Torsion and Gauge Invariance in Maxwell-Dirac Electrodynamics}

\author{H.T. Nieh}
\email{nieh@tsinghua.edu.cn}
\affiliation{Institute for Advanced Study, Tsinghua University, Beijing 100084, China}

\begin{abstract}
It has been known for a long time that the presence of torsion is in conflict with gauge invariance of the the electromagnetic field in curved Riemann-Cartan space if the Maxwell field is minimally coupled to the curved gravitational space through the covariant derivative. In search for a possible solution, we consider in this note the system of Maxwell-Dirac electrodynamics in Riemann-Cartan space. Through investigating consistency of the field equations, and taking cue from the scale invariance properties of the system, we come up with a solution that satisfies gauge invariance without having to dispense with torsion in the coupled Maxwell-Dirac system. This is achieved by modifying the connection that appears in the covariant derivative for the Maxwell field. The modified connection turns out to be in the form of a Weyl connection, with the torsion trace vector playing the effective role of a Weyl gauge field. With this modified connection, which is symmetrical, the Lorentz-spin current of the photon field is seen to vanish. In addition, except for the Dirac mass term, the system exhibits local scale invariance. The same consideration applies to all gauge theories, abelian or non-ableian, in the standard model of particle physics.
\end{abstract}


\date{January 27, 2017}
\maketitle

\section{Introduction}

Curvature and torsion are the two fundamental tensors in the Riemann-Cartan space. The basic variables describing the Riemann-Cartan space are the vierbein fields $e_{~\mu}^{a}$ and the Lorentz-spin connection fields $\omega_{~~\mu}^{ab}$, which, respectively, represent the translational gauge fields and the Lorentz-rotation gauge fields in the Einstein-Cartan-Sciama-Kibble theory of gravitation \cite{Kibble,Sciama}. While the energy-momentum tensor is defined as the response of the physical system to variations of the vierbein field, the Lorentz-spin current is defined as response to variations of the Lorentz-spin connection field \cite{Kibble,Sciama}. That torsion is not compatible with gauge invariance was already noted by Kibble \cite{Kibble} in his original paper, and discussed in the influential review paper of Hehl et al \cite{Hehl}. Gauge invariance can refer to the abelian U(1) gauge theory of electrodynamics as well as to the non-abelian gauge theories in the standard model of particle physics. This non-compatibility problem has since been discussed by various authors \cite{Benn,Sabbata,Andrade,Obukhov,Hammond,Itin}. One approach, and perhaps the consensus approach by now, is to postulate \cite{Hehl} that the Maxwell tensor $F_{\mu\nu}$ takes the form of the flat-space curl $A_{\mu,\nu}-A_{\nu,\mu}$, without, however, specifying the non-minimal covariant derivative that should be used to achieve this goal. Another is to bring into the system additional specific torsion sources, like in the work of Hojman et al \cite{Hojman}, which, in addition to being ad hoc and incompatible with experimental evidence \cite{Ni}, is likely to bring complications to renormalizability \cite{Panangaden,Lavrov} of the physical gauge theories in the standard model of particle physics. In this paper, we examine this question within the realistic physical system of Maxwell-Dirac electrodynamics. To achieve consistency of the field equations of the Maxwell-Dirac system, we shall see that the underlying connection for the Maxwell field can not be the symmetric Christoffel connection unless torsion also disappears in the Dirac sector, namely, torsion is completely dispensed with in the entire Maxwell-Dirac system. A clue in finding a potential candidate for a suitable connection to be adopted in the covariant derivative for the Maxwell field comes from the scale invariance properties of the field equation for the Dirac field, which explicitly show that the torsion trace vector effectively plays the role of the Weyl gauge field for local scale transformations \cite{Nieh}. This suggests that the connection should be modified so that the covariant derivative for the Maxwell field possesses transparent scale covariant transformation property. By imposing consistency of the field equations, the new connection for the Maxwell field turns out to be in the form of a Weyl connection, which is symmetric and compatible with gauge invariance. The resulting scheme is thus seen to possess not only gauge invariance but also desirable scale invariance properties.

\vspace{3mm}

\section{Maxwell-Dirac System in Riemann-Cartan Space}

We consider the Maxwell-Dirac system of electrodynamics in the background of curved Riemann-Cartan space, which is described by the vierbein field $e_{~\mu}^{a}$, their inverse $e_{a}^{~\mu}$, and the Lorentz-spin connection field $\omega_{~~\mu}^{ab}$. The metric is defined by
$$g_{\mu\nu}=\eta_{ab}e_{~\mu}^{a}e_{~\nu}^{b}, \eqno(1) $$

\noindent and the affine connection by
$$\Gamma_{~\mu\nu}^{\lambda}=e_{a}^{~\lambda}(a_{~\mu,\nu}^{a}+\omega_{~b\nu}^{a}e_{~\mu}^{b}), \eqno(2) $$

\noindent where $\eta_{ab}=(1,-1,-1,-1)$. The covariant derivatives with respect to both local Lorentz transformations and general coordinate transformations are defined, such as
$$\nabla_{\mu}\chi_{a}^{~\lambda}=\chi_{a~,\mu}^{~\lambda}-\omega_{~a\mu}^{b}\chi_{b}^{~\lambda}+\Gamma_{~\nu\mu}^{\lambda}\chi_{a}^{~\nu}, \eqno(3) $$
$$\nabla_{\mu}\chi_{~\nu}^{a}=\chi_{~\nu,\mu}^{a}+\omega_{~b\mu}^{a}\chi_{~\nu}^{b}-\Gamma_{~\nu\mu}^{\lambda}\chi_{~\lambda}^{a}. \eqno(4) $$
\noindent It can be easily verified that $\nabla_{\nu}e_{~\mu}^{a}=0$ and $\nabla_{\nu}e_{a}^{~\mu}=0$ so that
$$\nabla_{\lambda}g^{\mu\nu}=0, \eqno (5) $$
$$\nabla_{\lambda}g_{\mu\nu}=0. \eqno (6) $$

\noindent The connection $\Gamma_{~\mu\nu}^{\lambda}$ defined by (2) is thus metric-compatible. In general, it is not symmetric, and the anti-symmetric part is the torsion tensor:
$$C_{~\mu\nu}^{\lambda}=\Gamma_{~\mu\nu}^{\lambda}-\Gamma_{~\nu\mu}^{\lambda}. \eqno (7) $$

\noindent In the presence of torsion, the metric compatibility relations (5) and (6) imply that the connection is of the general form:
$$\Gamma_{~\mu\nu}^{\lambda}=\frac{1}{2}g^{\lambda\rho}(g_{\rho\mu,\nu}+g{\nu\rho,\mu}-g{\mu\nu,\rho})+Y_{~\mu\nu}^{\lambda}, \eqno (8) $$

\noindent where the contortion tensor $Y_{~\mu\nu}^{\lambda}$ is given by
$$Y_{~\mu\nu}^{\lambda}=\frac{1}{2}(C_{~\mu\nu}^{\lambda}+C_{\mu\nu}^{~~\lambda}+C_{\nu\mu}^{~~\lambda}). \eqno (9) $$

The basic field variables of the Maxwell-Dirac electrodynamics are the Maxwell field $A_{\mu}$ and Dirac field ${\psi}$. The action for the system is of the form
$$\begin{array}{rl}W=\displaystyle \int d^{4}x\epsilon
[-\frac{1}{4}F^{\mu\nu}F_{\mu\nu}\\
+\frac{1}{2}(\bar{\psi}i\gamma^{a}e_{a}^{~\mu}D_{\mu}\psi-\bar{\psi}\bar{D}_{\mu}i\gamma^{a}e_{a}^{~\mu}\psi)-m\bar{\psi}\psi], \end{array} \eqno (10) $$

\noindent where
$$D_{\mu}=\partial_{\mu}-\frac{i}{4}\sigma_{ab}\omega_{~~\mu}^{ab}, \eqno (11)  $$
$$\bar{D}_{\mu}=\bar{\partial}_{\mu}+\frac{i}{4}\sigma_{ab}\omega_{~~\mu}^{ab}, \eqno (12) $$

\noindent and $\epsilon=\det{e}_{~\mu}^{a}$. The partial $\bar{\partial}_{\mu}$ in (10) is understood to operate on $\bar{\psi}$ on the left, and $\sigma_{ab}=\frac{i}{2}[\gamma_{a},\gamma_{b}]$ \cite{Bjorken}. The Maxwell field strength $F_{\mu\nu}$ is defined as
$$F_{\mu\nu}=\nabla_{\mu}A_{\nu}-\nabla_{\nu}A_{\mu},$$

\noindent and $F^{\mu\nu}=g^{\mu\lambda}g^{\nu\rho}F_{\lambda\rho}$.

The Lagrangian in the action (10) is invariant under local Lorentz transformations, general coordinate transformations as well as local scale transformations (with m=0), the latter being defined, with the proper scale weights for the various fields, by
$$e_{a}^{~\mu}\rightarrow e^{-\Lambda(x)}e_{a}^{~\mu},$$
$$e_{~\mu}^{a}\rightarrow e^{\Lambda(x)}e_{~\mu}^{a},$$
$$\psi(x)\rightarrow e^{-\frac{3}{2}\Lambda(x)}\psi(x),$$
$$A_{\mu}(x)\rightarrow A_{\mu},$$
$$\omega_{~~\mu}^{ab}(x)\rightarrow \omega_{~~\mu}^{ab}(x).$$

\noindent However, the Lagrangian is not gauge invariant in the presence of torsion because, as is well known, $F_{\mu\nu}=A_{\nu,\mu}-A_{\mu,\nu}+C_{~\mu\nu}^{\lambda}A_{\lambda}$ is not.

The Euler-Lagrange equation for the Dirac field can be obtained straightforwardly. On account of
$$\epsilon^{-1}\epsilon_{,\mu}=\Gamma_{~\lambda\mu}^{\lambda}=\Gamma_{~\mu\lambda}^{\lambda}+C_{~\lambda\mu}^{\lambda}, \eqno (13)  $$

\noindent and the commutation properties of the Dirac gamma matrices \cite{Bjorken}, we obtain \cite{Nieh}
$$[i\gamma^{a}e_{a}^{\mu}(D_{\mu}+iA_{\mu}+\frac{1}{2}C_{~\lambda\mu}^{\lambda})-m]\psi=0, \eqno (14)  $$

\noindent where $D_{\mu}$ is given in (11). We know that the Lagrangian in the action (10) is scale invariant when m=0. The Dirac equation (14) is thus expected to be scale invariant except the mass term. We have, by its construction according to (2), the connection $\Gamma_{~\mu\nu}^{\lambda}$ has the following scale transformation property
$$\Gamma_{~\mu\nu}^{\lambda}\rightarrow \Gamma_{~\mu\nu}^{\lambda}+\delta_{~\mu}^{\lambda}\Lambda_{,\nu}, \eqno (15)  $$

\noindent which implies
$$C_{~\lambda\mu}^{\lambda}\rightarrow C_{~\lambda\mu}^{\lambda}+3\Lambda_{,\mu}. \eqno (16)  $$

\noindent We denote
$$B_{\mu}=\frac{1}{3}C_{~\lambda\mu}^{\lambda}. \eqno (17)  $$

\noindent It transforms as a Weyl gauge field for local scale transformations \cite{Nieh}
$$B_{\mu}\rightarrow B_{\mu}+\Lambda_{,\mu}. \eqno (18)  $$

\noindent The Dirac equation (13) is then expressed as
$$[i\gamma^{a}e_{a}^{\mu}(D_{\mu}+iA_{\mu}+\frac{3}{2}B_{\mu})-m]\psi=0. \eqno (19)  $$

\noindent So, indeed, except for the mass term, the Dirac equation written in this form shows explicit scale invariance, and with the proper scale weight $\frac{3}{2}$ for the Dirac field $\psi$.

The Euler-Lagrange equation for the Maxwell field is obtained straightforwardly. It is of the form
$$(\nabla_{\mu}+3B_{\mu})F^{\mu\nu}=J^{\mu}, \eqno (20)  $$

\noindent where the current $J_{\mu}$ is given by
$$J^{\mu}=\bar{\psi}\gamma^{a}e_{a}^{~\mu}\psi. \eqno (21)  $$

\section{Consistency of Field Equations}

In the presence of torsion, the field equation (20) is not gauge invariant. We would like to check whether current conservation is valid and whether the system of field equations, namely (19) and (20), are mutually consistent. As a consequence of the Dirac equation (19) and its hermitian conjugate equation for $\bar{\psi}$, it is straightforward to verify that the current $J^{\mu}$ is indeed conserved,
$$(\nabla_{\mu}+3B_{\mu})J^{\mu}=0. \eqno (22)  $$

\noindent Consistency of (20) with this current conservation equation (22) requires that
$$(\nabla_{\mu}+3B_{\mu})(\nabla_{\nu}+3B_{\nu})F^{\mu\nu}=0. \eqno (23)  $$

\noindent Making use of the anti-symmetry of $F^{\mu\nu}$, it is straightforward, though tedious, to show that
$$\begin{array}{rl}(\nabla_{\mu}+3B_{\mu})(\nabla_{\nu}+3B_{\nu})F^{\mu\nu}
=-R_{~\rho\mu\nu}^{\mu}F^{\rho\nu}\\+\frac{1}{2}C_{~\rho\nu}^{\mu}\nabla_{\mu}F^{\rho\nu}+\frac{3}{2}F^{\mu\nu}(\nabla_{\mu}B_{\nu}-\nabla_{\nu}B_{\mu}).\end{array}
\eqno (24)  $$

\noindent For the right-hand side of (24) to vanish, it is necessary, due to its structure, that the second term has to vanish. That is, we have to set $C_{~\rho\nu}^{\mu}=0$. When torsion vanishes, $B_{\mu}$ also vanishes, and the connection reduces to the Christoffel connection, implying that $R_{~\rho\mu\nu}^{\mu}$ is symmetric in $\rho$ and $\nu$. The three terms on the right-hand side of (24) thus all vanish. Consistency of the two field equations of the system (19) and (20) is seen to require the vanishing of torsion. And, as a result, the system becomes gauge invariant at the same time. However, the two field equations (19) and (20) are no longer scale invariant (with m=0), even though the action W in (10) remains invariant. It is also clear that if the Christoffel connection is adopted for the Maxwell sector, consistency requires that torsion is to be dropped from the coupled Dirac sector as well. That is, torsion is to be totally dispensed with in the system of Maxwell-Dirac electrodynamics. What we have learned here is that even though current conservation, being a consequence of the Dirac equation, does not depend on gauge invariance, consistency of the field equations of the coupled system does.

\section{Searching for Suitable Connection}

We have just seen that gauge invariance is indeed closely tied up with the consistency of the field equations. If, however, torsion is to play any role in the Maxwell-Dirac electrodynamics, we need to rescue it by finding a suitably modified connection that satisfies gauge invariance as well as achieves consistency of the field equations, without, however, requiring a vanishing torsion. A clue comes from observing the behavior of the field equations under scale transformations. We have noticed that the Dirac equation (19) has a clean scale transformation property. The torsion vector $B_{\mu}$ acts as an effective Weyl scale gauge field, with its coefficient in (19) properly reflecting the scale weight $\frac{3}{2}$ of the Dirac field $\psi$. If we look at the current conservation equation (22), we notice that the coefficient of the $B_{\mu}$ field is 3, though the current $J^{\mu}$ given by (21) actually carries a scale weight of 4, which includes the scale weight of $e_{a}^{~\mu}$. The mismatch is accounted for by the scale transformation property of the connection term in (22). Explicitly, (22) is
$$J_{~,\mu}^{\mu}+\Gamma_{~\nu\mu}^{\mu}J^{\nu}+3B_{\mu}J^{\mu}=0, \eqno (22')  $$

\noindent in which the connection transforms, according to (15), as
$$\Gamma_{~\nu\mu}^{\mu}\rightarrow \Gamma_{~\nu\mu}^{\mu}+\Lambda_{,\nu},  $$

\noindent which makes up for the missing scale weight. This suggests the consideration of a modified connection
$$\tilde{\Gamma}_{~\nu\mu}^{\lambda}=\Gamma_{~\nu\mu}^{\lambda}-\delta_{~\nu}^{\lambda}B_{\mu}, \eqno (25)  $$

\noindent which is invariant under scale transformations. In terms of the newly defined connection, current conservation (22') takes the form
$$J_{~,\mu}^{\mu}+\tilde{\Gamma}_{~\nu\mu}^{\mu}J^{\nu}+4B_{\mu}J^{\mu}=0, \eqno (26')  $$

\noindent or, equivalently,
$$(\tilde{\nabla}_{\mu}+4B_{\mu})J^{\mu}=0, \eqno (26)  $$

\noindent which properly accounts for the scale weight 4 of the current $J^{\mu}$, recalling that $e_{a}^{\mu}$ in $J^{\mu}$ has a weight of 1.

In the Maxwell equation (20), similarly, there is an apparent mismatch of the scale dimensions. Corresponding to the newly defined connection, the Maxwell tensor is defined as
$$\tilde{F}_{\mu\nu}=\tilde{\nabla}_{\mu}A_{\nu}-\tilde{\nabla}_{\nu}A_{\mu}, \eqno (27)  $$

\noindent It is this newly defined $\tilde{F}_{\mu\nu}$ that substitutes $F_{\mu\nu}$ in the action W in (10). While the resulting field equation for the Dirac field stays unchanged as in (19), the corresponding Euler-Lagrange equation for the Maxwell field takes the form
$$(\tilde{\nabla}_{\nu}+4B_{\nu})\tilde{F}^{\mu\nu}=J^{\mu}, \eqno (28)  $$

\noindent which properly reflects the scale weight 4 of $\tilde{F}^{\mu\nu}$.

\section{Gauge Invariance and Scale Invariance}

With the Maxwell-Dirac system treated with the newly defined connection, we again check the consistency of the corresponding field equations (19) and (28). Consistency requires that (19) is compatible with the current conservation (26), namely,
$$(\tilde{\nabla}_{\mu}+4B_{\mu})(\tilde{\nabla}_{\nu}+4B_{\nu})\tilde{F}^{\mu\nu}=0. \eqno (29)  $$

\noindent The left-hand side of the equation can be straightforwardly calculated. The result is
$$-\tilde{R}_{~\rho\mu\nu}^{\mu}\tilde{F}^{\rho\nu}+\frac{1}{2}\tilde{C}_{~\rho\nu}^{\mu}\tilde{\nabla}_{\mu}\tilde{F}^{\rho\nu}+2\tilde{F}^{\mu\nu}(\nabla_{\mu}B_{\nu}-\nabla_{\nu}B_{\mu}). \eqno (30)  $$

\noindent For this to vanish, it is necessary that
$$\tilde{C}_{~\rho\nu}^{\mu}=\tilde{\Gamma}_{~\rho\nu}^{\mu}-\tilde{\Gamma}_{~\nu\rho}^{\mu}=0, \eqno (31)  $$
\noindent which means that $\tilde{\Gamma}_{~\mu\nu}^{\lambda}$ is symmetric and implies that
$$C_{~\mu\nu}^{\lambda}=g_{~\mu}^{\lambda}B_{\nu}-g_{~\nu}^{\lambda}B_{\mu}. \eqno(32)  $$

\noindent The contortion tensor $Y_{~\mu\nu}^{\lambda}$ is calculated according to (9) to yield
$$Y_{~\mu\nu}^{\lambda}=g_{\mu\nu}B^{\lambda}-g_{~\nu}^{\lambda}B_{\mu}. \eqno (33)  $$

\noindent We then obtain from (8) and (25) the results
$$\Gamma_{~\mu\nu}^{\lambda}=\frac{1}{2}g^{\lambda\rho}(g_{\rho\mu,\nu}+g_{\nu\rho,\mu}-g_{\mu\nu,\rho})+g_{\mu\nu}B^{\lambda}-g_{~\nu}B_{\mu},  \eqno (34)$$
$$\tilde{\Gamma}_{~\mu\nu}^{\lambda}=\frac{1}{2}g^{\lambda\rho}(g_{\rho\mu,\nu}+g_{\nu\rho,\mu}-g_{\mu\nu,\rho})+g_{\mu\nu} B^{\lambda}-g_{~\mu}^{\lambda}B_{\nu}-g_{~\nu}^{\lambda}B_{\mu}. \eqno (35)   $$

With the result (35), we calculate the antisymmetric part of $\tilde{R}_{~\rho\mu\nu}^{\mu}$ in (30) and obtain
$$\frac{1}{2}(\tilde{R}_{~\rho\mu\nu}^{\mu}-\tilde{R}_{~\nu\mu\rho}^{\mu})=2(B_{\nu,\rho}-B_{\rho,\nu}).  $$

\noindent The first term and third term in (30) exactly cancel each other out. Indeed, with the modified connection $\tilde{\Gamma}_{~\mu\nu}^{\lambda}$ given in (35), consistency of the field equations (19) and (28) is established. At the same time, $\tilde{\Gamma}_{~\mu\nu}^{\lambda}$ being symmetric, $\tilde{F}_{\mu\nu}=A_{\nu,\mu}-A_{\mu,\nu}$, gauge invariance is also restored.

The Affine connection $\tilde{\Gamma}_{~\mu\nu}^{\lambda}$ given in (35) is seen to have exactly the same structure as a Weyl connection, with the torsion trace vector $B_{\mu}=\frac{1}{3}C_{~\lambda\mu}^{\lambda}$, a geometric entity in the Riemann-Cartan space, effectively playing the role of the scale gauge field. With this connection adopted for the Maxwell-Dirac electrodynamics, the system is self consistent, gauge invariant and scale invariant (with m=0), without having to dispense with torsion. In contrast to the purely metric Christoffel connection, the incorporation of torsion into the Weyl-like connection enables the system to exhibit good scale transformation properties, in addition to maintaining gauge invariance.

\section{Spin Current and Torsion}

The relationship between torsion and spin is an essential feature of the Einstein-Cartan-Sciama-Kibble theory of gravitation. The spin current $S_{ab}^{~~\mu}$ is defined as the functional derivative of the action W with respect to the spin-connection $\omega_{~~\mu}^{ab}$:
$$\delta W=\int d^{4}xh\frac{1}{2}S_{ab}^{~~\mu}\delta\omega_{~~\mu}^{ab}.  $$

\noindent The contribution to the spin current by the Dirac field is well-known \cite{Kibble,Hehl,Nieh} and given by
$$\frac{1}{2}\varepsilon_{abcd}\bar{\psi}\gamma_{5}\gamma^{d}\psi.   $$

\noindent The contribution from the Maxwell photon field can be obtained by varying $-\frac{1}{4}\tilde{F}^{\mu\nu}\tilde{F}_{\mu\nu}$ in the action W with respect to $\omega_{~~\mu}^{ab}$. It vanishes identically, since the symmetric connection $\tilde{\Gamma}_{~\nu\mu}^{\lambda}$ completely drops out from $\tilde{F}^{\mu\nu}$. Thus, the Maxwell photon field has no contribution to the spin current, and, as a result, neither to torsion. It is clear that gauge invariance of the Maxwell tensor $F_{\mu\nu}$ requires a symmetric connection, which, in turn, implies that photon does not give rise to torsion.

\section{Concluding Remarks}

Ever since the advent of the Einstein-Cartan-Sciama-Kibble theory of gravitation, compatibility of torsion with gauge invariance has been a problem. Gauge invariance can refer to abelian U(1) gauge theories as well as to non-abelian gauge theories in the standard model of particle physics. The compatibility problem would be an obstacle for Cartan's torsion to play a role in the real world of physics. As such, it is a hurdle waiting to be overcome. To maintain gauge invariance, we do need a symmetric connection for the Maxwell field. But the symmetric Christoffel connection is not the answer, as its adoption would imply that torsion should vanish in the coupled Dirac sector as well, and the end result would be that torsion is absent in the entire coupled Maxwell-Dirac system. We have come up with a solution that exhibits transparent gauge invariance, and, in addition, scale invariance in the system (except for the Dirac mass term), without dispense with torsion. This same solution applies as well to other gauge theories, abelian or non-abelian. The solution is embodied in a modified affine connection that possesses the structure of a Weyl connection, with the torsion trace vector effectively serving the role of the Weyl scale gauge potential. For any given Riemann-Cartan background, torsion is coupled to the Maxwell field $A_{\mu}$ through this connection, though it disappears in $F_{\mu\nu}$. Torsion is still effectively disengaged from the photons, but stays engaged with the coupled Dirac field. This connection represents perhaps a minimal extension of the "minimal coupling" procedure for photons in the general Riemann-Cartan space. In a sense, it provides an explicit and  specific justification for the ansatz of Hehl et al \cite{Hehl} in taking the Maxwell field strength in the form of its flat-space curl, without having to dispense with torsion.

\vspace{2mm}
\begin{acknowledgments}

The author would like to thank Professor Friedrich Hehl for his thoughtful comments.

\end{acknowledgments}

\end{document}